\documentclass{article}
\usepackage{amsmath,cite}
\usepackage{graphicx}
\usepackage{color}
\usepackage{amsmath, amssymb, amsfonts, amsthm}
\usepackage{algorithm}
\usepackage{algpseudocode}
\usepackage{subfigure}
\usepackage[hyphens]{url}
\usepackage[margin=1in]{geometry}

\newtheorem{definition}{Definition}

\usepackage{mathtools}

\title{Cover Song Identification with Timbral Shape Sequences}

\author{ Christopher J. Tralie \\ 
        \texttt{chris.tralie@gmail.com}
        \and
        Paul Bendich\\
        \texttt{bendich@math.duke.edu} 
}

\begin{document}
\maketitle
%


\begin{abstract}
We introduce a novel low level feature for identifying cover songs which quantifies the relative changes in the smoothed frequency spectrum of a song.  Our key insight is that a sliding window representation of a chunk of audio can be viewed as a time-ordered point cloud in high dimensions.  For corresponding chunks of audio between different versions of the same song, these point clouds are approximately rotated, translated, and scaled copies of each other.  If we treat MFCC embeddings as point clouds and cast the problem as a relative shape sequence, we are able to correctly identify 42/80 cover songs in the ``Covers 80" dataset.  By contrast, all other work to date on cover songs exclusively relies on matching note sequences from Chroma derived features.
\end{abstract}
\section{Introduction}\label{sec:introduction}

Automatic cover song identification is a surprisingly difficult classical problem that has long been of interest to the music information retrieval community \cite{ellis2006identifying}.  
This problem is significantly more challenging than traditional audio fingerprinting because a combination of tempo changes, musical key transpositions, embellishments in time and expression, and changes in vocals and instrumentation can all occur simultaneously between the original version of a song and its cover. Hence, low level features used in this task need to be robust to all of these phenomena, ruling out raw forms of popular features such as MFCC, CQT, and Chroma.

One prior approach, as reviewed in Section~\ref{sec:priorwork}, is to compare beat-synchronous sequences of chroma vectors between candidate covers. The beat-syncing helps this be invariant to tempo, but it is still not invariant to key.  However, many schemes have been proposed to deal with this, up to and including a brute force check over all key transpositions.

Chroma representations factor out some timbral information by folding together all octaves, which is sensible given the effect that different instruments and recording environments have on timbre.  However, valuable non-pitch information which is preserved between cover versions, such as spectral fingerprints from drum patterns, is obscured in Chroma representation.  This motivated us to take another look at whether timbral-based features could be used at all for this problem.  Our idea is that even if absolute timbral information is vastly different between two versions of the same song, the {\em relative evolution} of timbre over time should be comparable.

With careful centering and normalization within small windows to combat differences in global timbral drift between the two songs, we are indeed able to design shape features which are approximately invariant to cover. These features, which are based on self-similarity matrices of MFCC coefficients, can be used on their own to effectively score cover songs.  This, in turn, demonstrates that even if absolute pitch is obscured and blurred, cover song identification is still possible.

 Section~\ref{sec:priorwork} reviews prior work in cover song identification. Our method is described in detail by Sections~\ref{sec:blockpc} and ~\ref{sec:globalcompare}.  Finally, we report results on the ``Covers 80" benchmark dataset \cite{ellis2007covers80} in Section~\ref{sec:Results}, and we apply our algorithm to the recent ``Blurred Lines" copyright controversy.

\section{Prior Work}
\label{sec:priorwork}
To the best of our knowledge, all prior low level feature design for cover song identification has focused on Chroma-based representations alone.  The cover songs problem statement began with the work of \cite{ellis2006identifying}, which used FFT-based cross-correlation of all key transpositions of beat-synchronous chroma between two songs.  A follow-up work \cite{ellis2007} showed that high passing such cross-correlation can lead to better results.  In general, however, cross-correlation is not robust to changes in timing, and it is also a global alignment technique.  Serra \cite{serra2007music} extended this initial work by considering dynamic programming local alignment of chroma sequences, with follow-up work and rigorous parameter testing and an ``optimal key transposition index" estimation presented in \cite{serra2008chroma}.  The same authors also showed that a delay embedding of statistics spanning multiple beats before local alignment improves classification accuracy \cite{serra2009cross}.  In a different approach, \cite{kim2008music} compared modeled covariance statistics of all chroma bins, as well as comparing covariance statistics for all pairwise differences of beat-level chroma features, which is not unlike the ``bag of words" and bigram representations, respectively, in text analysis.  Other work tried to model sequences of chords \cite{bello2007audio} as a slightly higher level feature than chroma.  Slightly later work concentrated on fusing the results of music separated into melody and accompaniment \cite{foucard2010multimodal} and melody, bass line, and harmony \cite{salamon2012melody}, showing improvements over matching chroma on the raw audio.  The most recent work on cover song identification has focused on fast techniques for large scale pitch-based cover song identification, using a sparse set of approximate nearest neighbors \cite{tavenard2012efficient} and low dimensional projections \cite{humphrey2013data}.  Authors in \cite{ellis2012large} and \cite{nieto2014music} also use the {\em magnitude} of the 2D Fourier Transform of a sequences of chroma vectors treated as an image, so the resulting coefficients will be automatically invariant to key and time shifting without any extra computation, at the cost of some discriminative power.

Outside of cover song identification, there are other works which examine gappy sequences of MFCC in music, such as \cite{casey2006importance}.  However, these works look at matched sequences of MFCC-like features in their original feature space.  By contrast, in our work, we examine the {\em relative} shape of such features.  Finally, we are not the first to consider shape in an applied musical context.  For instance, \cite{urbano2011melodic} turns sequences of notes in sheet music into plane curves, whose curvature is then examined.  To our knowledge, however, we are the first to explicitly model shape in musical audio for version identification.

\section{Time Ordered Point Clouds from Blocks of Audio}
\label{sec:blockpc}

The first step of our algorithm uses a timbre-based method to turn a block of audio
into what we call a \emph{time-ordered point cloud.}  We can then compare to other time-ordered point clouds in a rotation, translation, and scale invariant manner using normalized Euclidean Self-Similarity matrices (Section~\ref{sec:SSMs}).  The goal is to then match up the relative shape of musical trajectories between cover versions.

\subsection{Point Clouds from Blocks and Windows}
\label{sec:SlidingWindows}

We start with a song, which is a function of time $f(t)$ that has been discretized as some vector $X$.  In the following discussion, the symbol $X(a, b)$ means the song portion beginning at time $t = a$ and ending at time $t = b$.  Given $X$, there are many ways to summarize a chunk of audio $w \in X$, which we call a {\em window}, as a point in some feature space.  We use the classical Mel-Frequency Cepstral coefficient representation \cite{Bogert1963}, which is based on a perceptually motivated log frequency and log power short-time Fourier transform that preserves timbral information.  In our application, we perform an MFCC with 20 coefficients, giving rise to a 20-dimensional point.
\begin{equation}
MFCC(w) \in \mathbb{R}^{20}
\end{equation}

Given a longer chunk of audio, which we call a {\em block}, we can use the above embedding on a collection of $K$ windows that cover the block to construct a collection of points, or a {\em point cloud}, representing that block.  More formally, given a block covering a range $[t_1, t_2]$, we want a set of window intervals $[a_i, b_i]$, with $i = 1 .. K$, so that
\begin{itemize}
\item $a_i < b_i$
\item $a_i < a_{i+1}$, $b_i < b_{i+1}$
\item $\cup_{i = 1}^K [a_i, b_i] = [t_1, t_2]$
\end{itemize}

Where $t_1$, $t_2$, $a_i$, and $b_i$ are all discrete time indices into the sampled audio $X$.  Hence, our final operator takes a set of time-ordered intervals $\{[a_1, b_1], [a_2, b_2], ..., [a_K, b_K]\}$ which cover a block $[t_1, t_2]$ and turns them into a $K$-dimensional point cloud in $\mathbb{R}^{20}$
\begin{equation}
\begin{split}
PC(\{[a_1, b_1], ..., [a_K, b_K]\}) = \\
  \{MFCC(X(a_1, b_1)), ..., MFCC(X(a_K, b_K))\}
\end{split}
\end{equation}
\subsection{Beat-Synchronous Blocks}

As many others in the MIR community have done, including \cite{ellis2006identifying} and \cite{ellis2007} for the cover songs application, we compute our features synchronized within beat intervals.  We use a simple dynamic programming beat tracker developed in \cite{ellis2007beat}.  Similarly to \cite{ellis2007}, we bias the beat tracker with three initial tempo levels: 60BPM, 120BPM, and 180BPM, and we compare the embeddings from all three levels against each other when comparing two songs, taking the best score out of the 9 combinations.  This is to mitigate the tendency of the beat tracker to double or halve the true beat intervals of different versions of the same song when there are tempo changes between the two.  The trade-off is of course additional computation.  We should note that other cover song works, such as \cite{serra2008chroma}, avoid beat tracking step altogether, hence bypassing these problems.  However, it is important for us to align our sequences as well as possible in time so that shape features are in correspondence, and this is a straightforward way to do so.

Given a set of beat intervals, the union of which makes up the entire song, we take blocks to be all contiguous groups of $B$ beat intervals.  In other words, we create a sequence of overlapping blocks $X_1, X_2, ...$ such that $X_i$ is made up of $B$ time-contiguous beat intervals, and $X_i$ and $X_{i+1}$ differ only by the starting beat of $X_i$ and the finishing beat of $X_{i+1}$.  Hence, given $N$ beat intervals, there are $N - B + 1$ blocks total.  Note that computing an embedding over more than one beat is similar in spirit to the chroma delay embedding approach in \cite{serra2009cross}.  Intuitively, examining patterns over a group of beats gives more information than one beat alone, the effect of which is empirically evaluated in Section~\ref{sec:Results}.  For all blocks, we take the window size $W$ to be the length of the average tempo period, and we advance the window intervals evenly from the beginning of the block to the end of a block with a {\em hop size} $H = W/200$.  Hence, there is a $99.5 \%$ overlap between windows.  We were inspired by theory on raw 1D time series signals \cite{perea2013sliding}, which shows that matching the window length to be just under the length of the period in a delay embedding maximizes the roundness of the embedding.  Here we would like to match beat-level periodicities and fluctuations therein, so it is sensible to choose a window size corresponding to the tempo.  This is in contrast to most other applications that use MFCC sliding window embeddings, which use a much smaller window size on the order of 10s of milliseconds, generally with a $50 \%$ overlap, to ensure that the frequency statistics are stationary in each window.  In our application, however, we have found that a longer window size makes our self similarity matrices (Section~\ref{sec:SSMs}) smoother, allowing for more reliable matches of beat-level musical trajectories, while having more windows per beat (high overlap) leads to more robust matching of SSMs using L2 (Section~\ref{sec:LocalAlignment}).
  
\begin{figure}
	\centering
	\subfigure[Window size 0.05 seconds]{  \includegraphics[width=0.46\columnwidth]{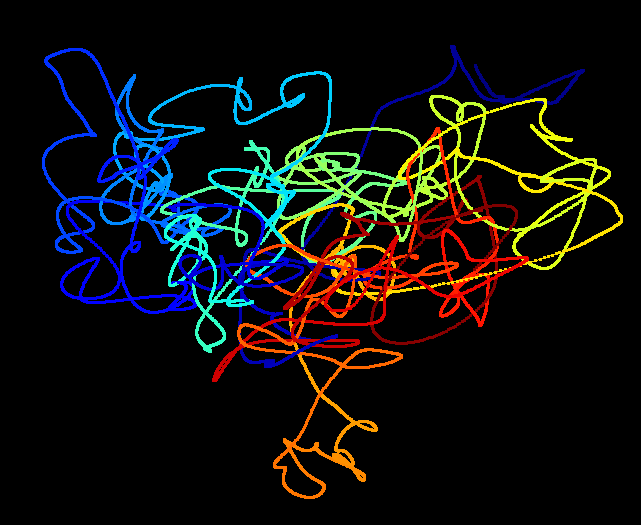} }%
	\subfigure[Window size 0.5 seconds]{  \includegraphics[width=0.46\columnwidth]{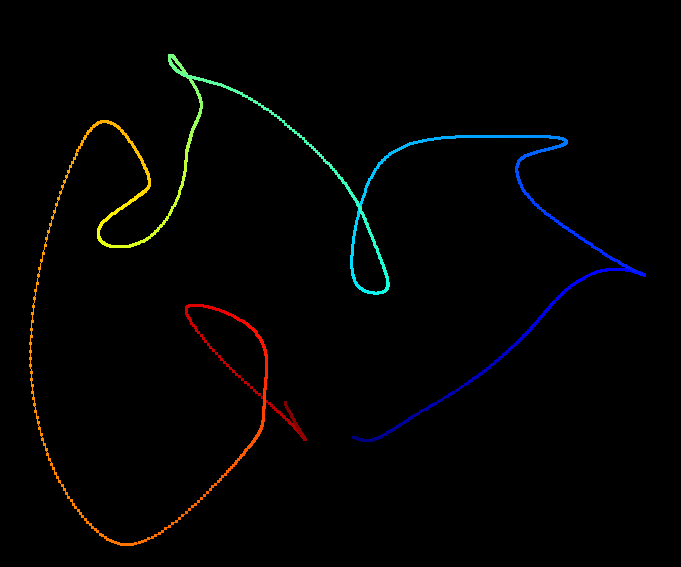} }	
	\caption{A screenshot from our GUI showing PCA on the sliding window representation of an 8-beat block from the hook of Robert Palmer's ``Addicted To Love" with two different window sizes.  Cool colors indicate windows towards the beginning of the block, and hot colors indicate windows towards the end.}
	\label{fig:PCAMFCC}
\end{figure}

Figure~\ref{fig:PCAMFCC} shows the first three principal components of an MFCC embedding with a traditional small window size versus our longer window embedding to show the smoothing effect.

\subsection{Euclidean Self-Similarity Matrices}
\label{sec:SSMs}

For each beat-synchronous block $X_l$ spanning $B$ beats, we have a 20-dimensional point cloud extracted from the sliding window MFCC representation.  Given such a time-ordered point cloud, there is a natural way to create an image which represents the shape of this point cloud in a rotation and translation invariant way, called the {\em self-similarity matrix} (SSM) representation.

\begin{figure}
	\centering
	\subfigure[A block of 4 beats with 400 windows sliding in the song ``We Can Work It Out" by The Beatles with a cover by Five Man Acoustical Jam]{  \includegraphics[width=\columnwidth]{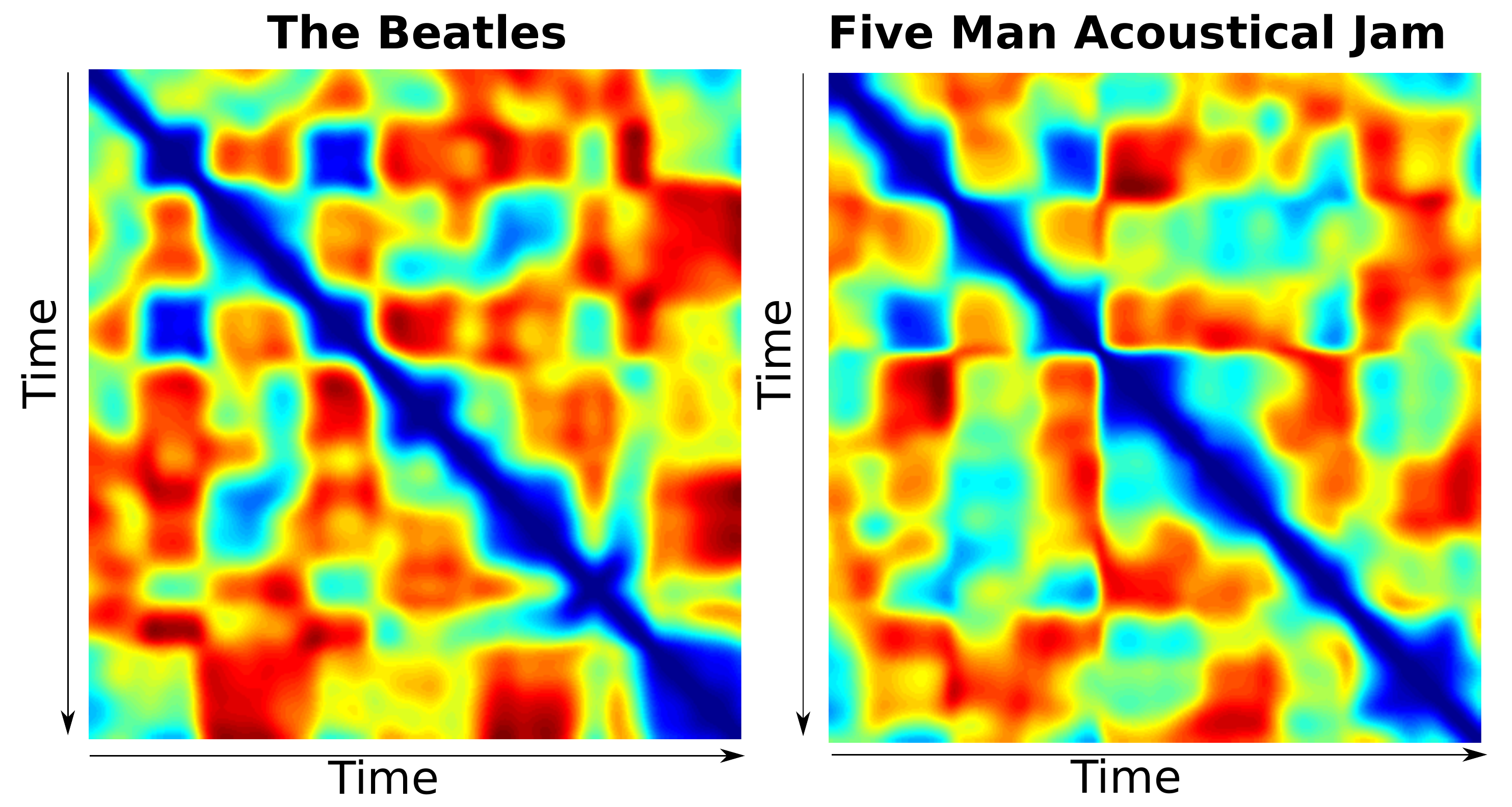} }	
	\subfigure[A block of 4 beats with 400 windows sliding in the song ``Don't Let It Bring You Down" by Neil Young with a cover by Annie Lennox.]{  \includegraphics[width=\columnwidth]{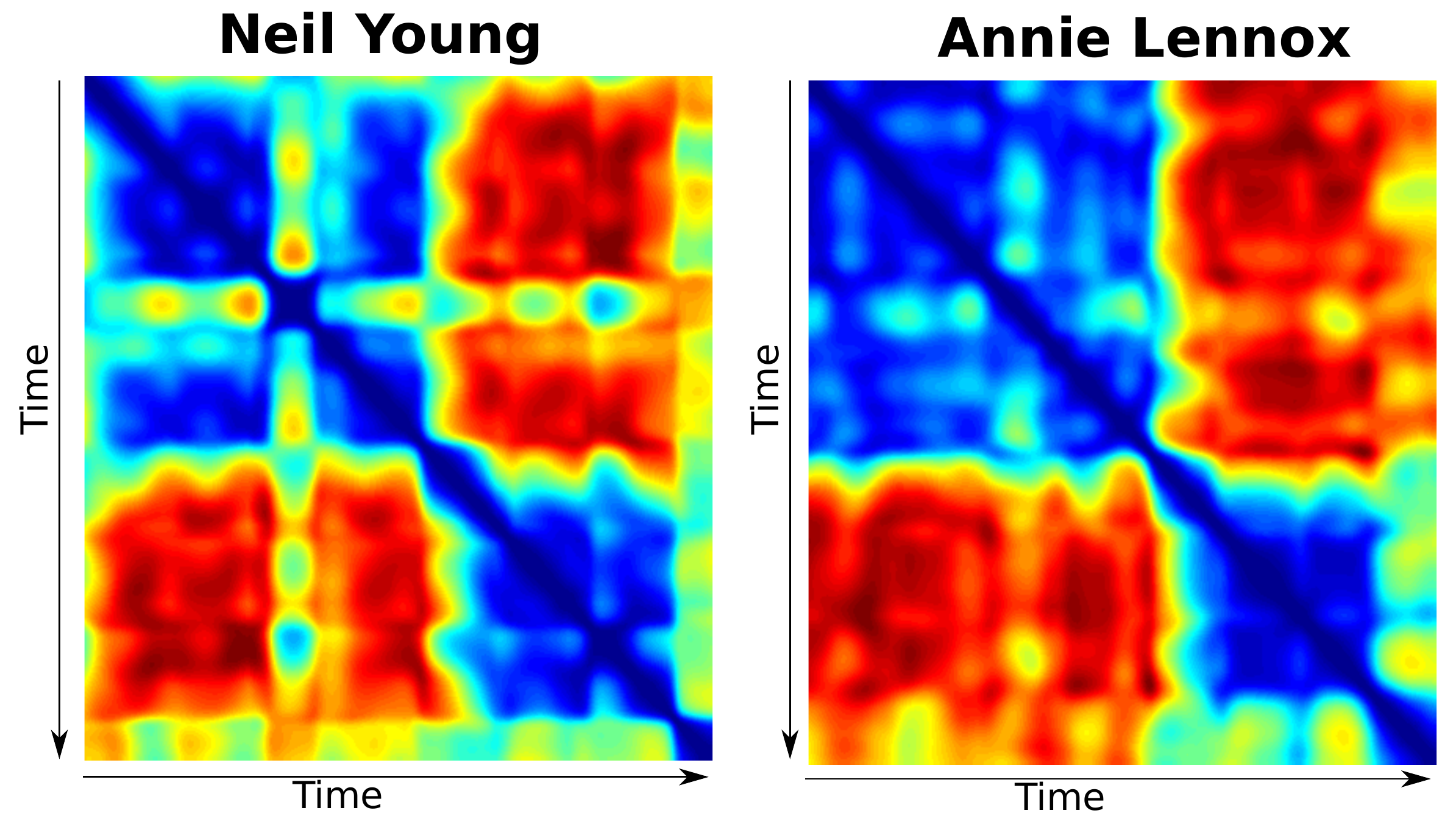} }%
	\caption{Two examples of MFCC SSM blocks which were matched between a song and its cover in the covers80 dataset.  Hot colors indicate windows in the block are far from each other, and cool colors indicate that they are close.}
	\label{fig:MFCCSSMExamples}
\end{figure}

\begin{definition}

A Euclidean {\em Self-Similarity Matrix (SSM)} over an ordered point cloud $X_l \in \mathbb{R}^{M \times k}$ is an $M \times M$ matrix $D$ so that
\begin{equation}
D_{ij} = ||X_l[i] - X_l[j]||_2
\end{equation}

\end{definition}

In other words, an SMM is an image representing all pairwise distances between points in a point cloud ordered by time.  SSMs have been used extensively in the MIR community already, spearheaded by the work of Foote in 2000 for note segmentation in time \cite{foote2000automatic}.  They are now often used in general segmentation tasks \cite{serra2012unsupervised} \cite{mcfee2014analyzing}.  They have also been successfully applied in other communities, such as computer vision to recognize activity classes in videos from different points of view and by different actors \cite{junejo2008cross}.  Inspired by this work, we use self-similarity matrices as isometry invariant descriptors of local shape in our sliding windows of beat blocks, with the goal of capturing relative shape.  In our case, the ``activities" are musical expressions over small intervals, and the ``actors" are different performers or groups of instruments.

To help normalize for loudness and other changes in relationships between instruments, we first center the point cloud within each block on its mean and scale each point to have unit norm before computing the SSM.  That is, we compute the SSM on $\hat{X^l}$, where

\begin{equation}
\hat{X_l} = \left\{ \frac{x - mean(x)}{||x - mean(x)||_2} : x \in X_l \right\}
\end{equation}

Also, not every beat block has the same number of samples due to natural variations of tempo in real songs.  Thus, to allow comparisons between all blocks, we resize each SSM to a common image dimension $d \times d$, which is a parameter chosen in advance, the effects of which are explored empirically in Section~\ref{sec:Results}.

Figure~\ref{fig:MFCCSSMExamples} shows examples of SSMs of 4-beat blocks pulled from the Covers80 dataset that our algorithm matches between two different versions of the same song.  Visually, similarities in the matched regions are evident.  In particular, viewing the images as height functions, many of the critical points are close to each other.  The ``We Can Work It Out" example shows how this can work even for live performances, where the overall acoustics are quite different.  Even more strikingly, the ``Don't Let It Bring You Down" example shows how similar shape patterns emerge even with an opposite gender singer and radically different instrumentation.  Of course, in both examples, there are subtle differences due to embellishments, local time stretching, and imperfect normalization between the different versions, but as we show in Section~\ref{sec:Results}, there are often enough similarities to match up blocks correctly in practice.

\section{Global Comparison of Two Songs}
\label{sec:globalcompare}

\begin{figure*}
	\centering
	\includegraphics[width=\textwidth]{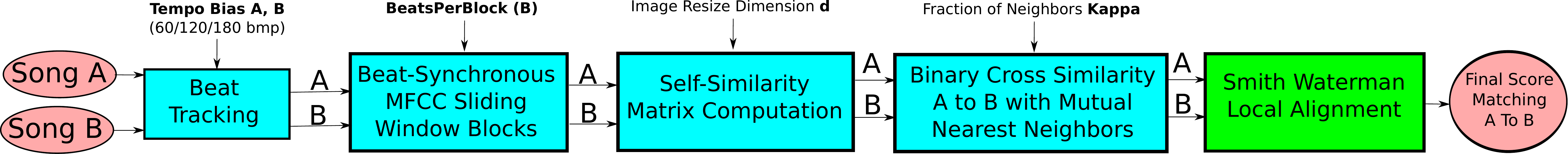}%
	\caption{A block diagram of our system for computing a cover song similarity score of two songs using timbral features.}
	\label{fig:BlockDGM}
\end{figure*}

Once all of the beat-synchronous SSMs have been extracted from two songs, we do a global comparison between all SSMs from two songs to score them as cover matches.  Figure~\ref{fig:BlockDGM} shows a block diagram of our system.  After extracting beat-synchronous timbral shape features on SSMs, we then extract a binary cross-similarity matrix based on the L2 distance between all pairs of self-similarity matrices between two songs.  We subsequently apply the Smith Waterman algorithm on the binary cross-similarity matrix to score a match between the two songs.  

\subsection{Binary Cross-Similarity And Local Alignment}
\label{sec:LocalAlignment}

Given a set of $N$ beat-synchronous block SSMs for a song A and a set of $M$ beat-synchronous block SSMs for a song B, we compute a song-level matching between song A and B by comparing all pairs of SSMs between the two songs.  For this we create an $N \times M$ cross-similarity matrix (CSM), where
\begin{equation}
\text{CSM}_{ij} = ||\text{SSMA}_i - \text{SSMB}_j||_2
\end{equation}
is the Frobenius norm (L2 image norm) between the SSM for the $i^{\text{th}}$ beat block from song A and the SSM for $j^{\text{th}}$ beat block for song B.  Given this cross-similarity information, we then compute a binary cross similarity matrix $B^M$.  A binary matrix is necessary so that we can apply the Smith Waterman local alignment algorithm \cite{smith1981identification} to score the match between song A and B, since Smith Waterman only works on a discrete, quantized alphabet, not real values \cite{serra2008chroma}.  To compute $B^M$, we take the mutual fraction $\kappa$ nearest neighbors between song A and song B, as in \cite{serra2009cross}.  That is, $B^M_{ij} = 1$ if $CSM_{ij}$ is within the $\kappa M^{\text{th}}$ smallest values in row $i$ of the CSM and if $CSM_{ij}$ is within the $\kappa N^{\text{th}}$ smallest values in column $j$ of the CSM, and 0 otherwise.  As in \cite{serra2009cross}, we found that a dynamic distance threshold for mutual nearest neighbors per element worked significantly better than a fixed distance threshold for the entire matrix.

\begin{figure}
	\centering
	\subfigure[Full cross-similarity matrix (CSM)]{  \includegraphics[width=0.46\columnwidth]{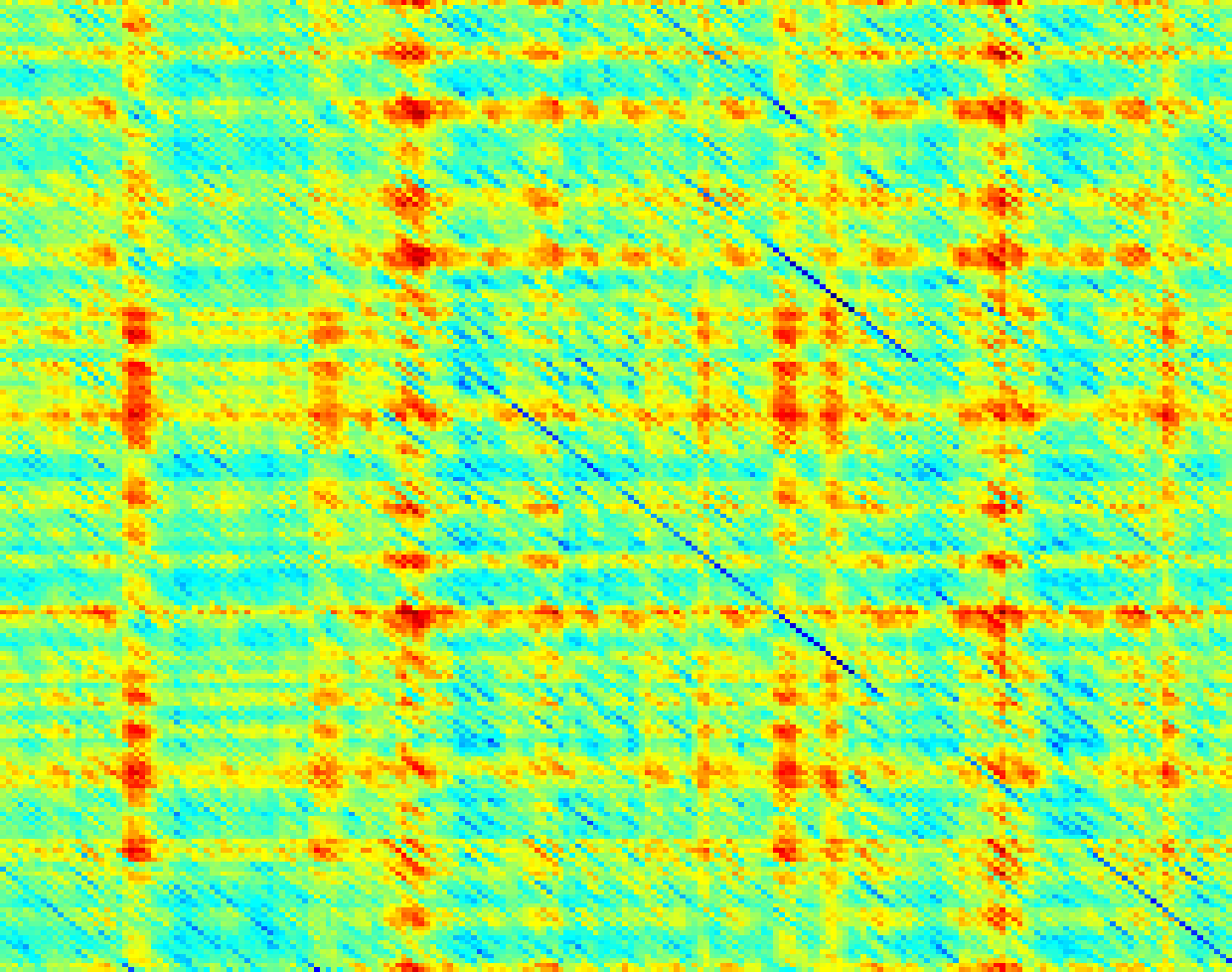} }
	\subfigure[$212 \times 212$ Binary cross-similarity matrix ($B^M$) with $\kappa = 0.05$]{  \includegraphics[width=0.46\columnwidth]{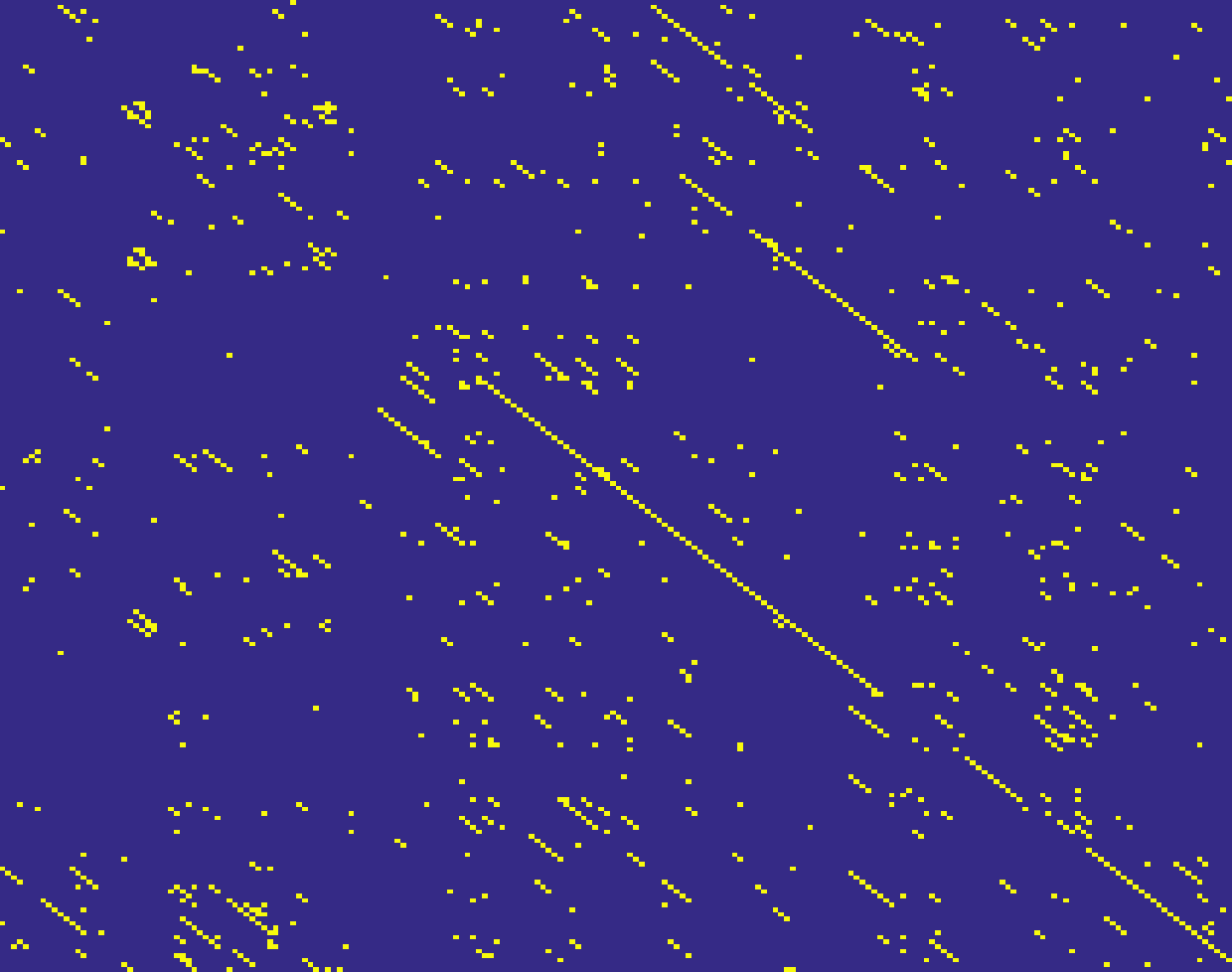} }
	\subfigure[Smith Waterman with local constraints: Score 93.1 ]{  \includegraphics[width=0.46\columnwidth]{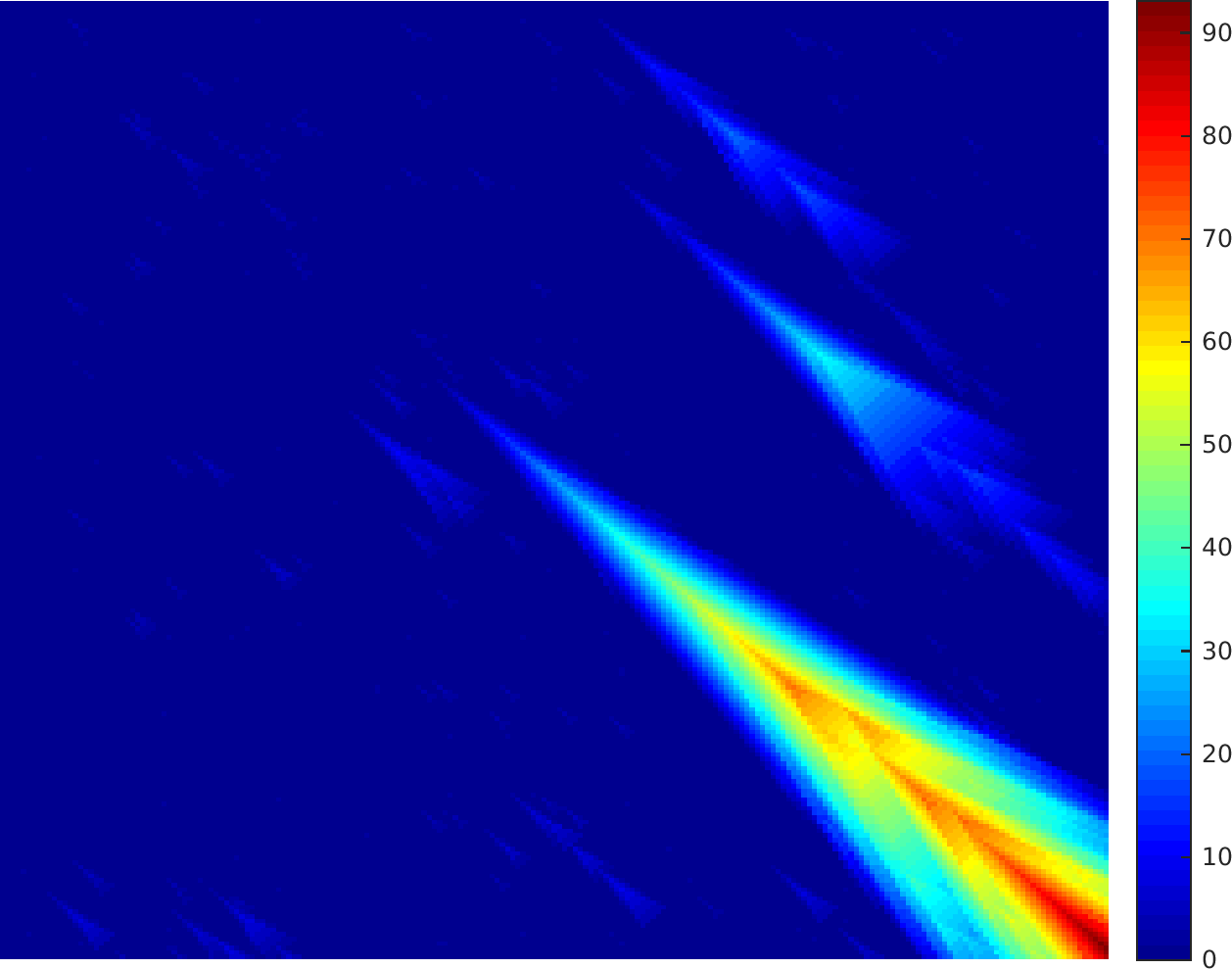} }
	\caption{Cross-similarity matrix and Smith Waterman on MFCC-based SSMs for a true cover song pair of ``We Can Work It Out" by The Beatles and Five Man Acoustical Jam.}
	\label{fig:CrossSimilarityMatch}
\end{figure}

\begin{figure}
	\centering
	\subfigure[Full cross-similarity matrix (CSM)]{  \includegraphics[width=0.46\columnwidth]{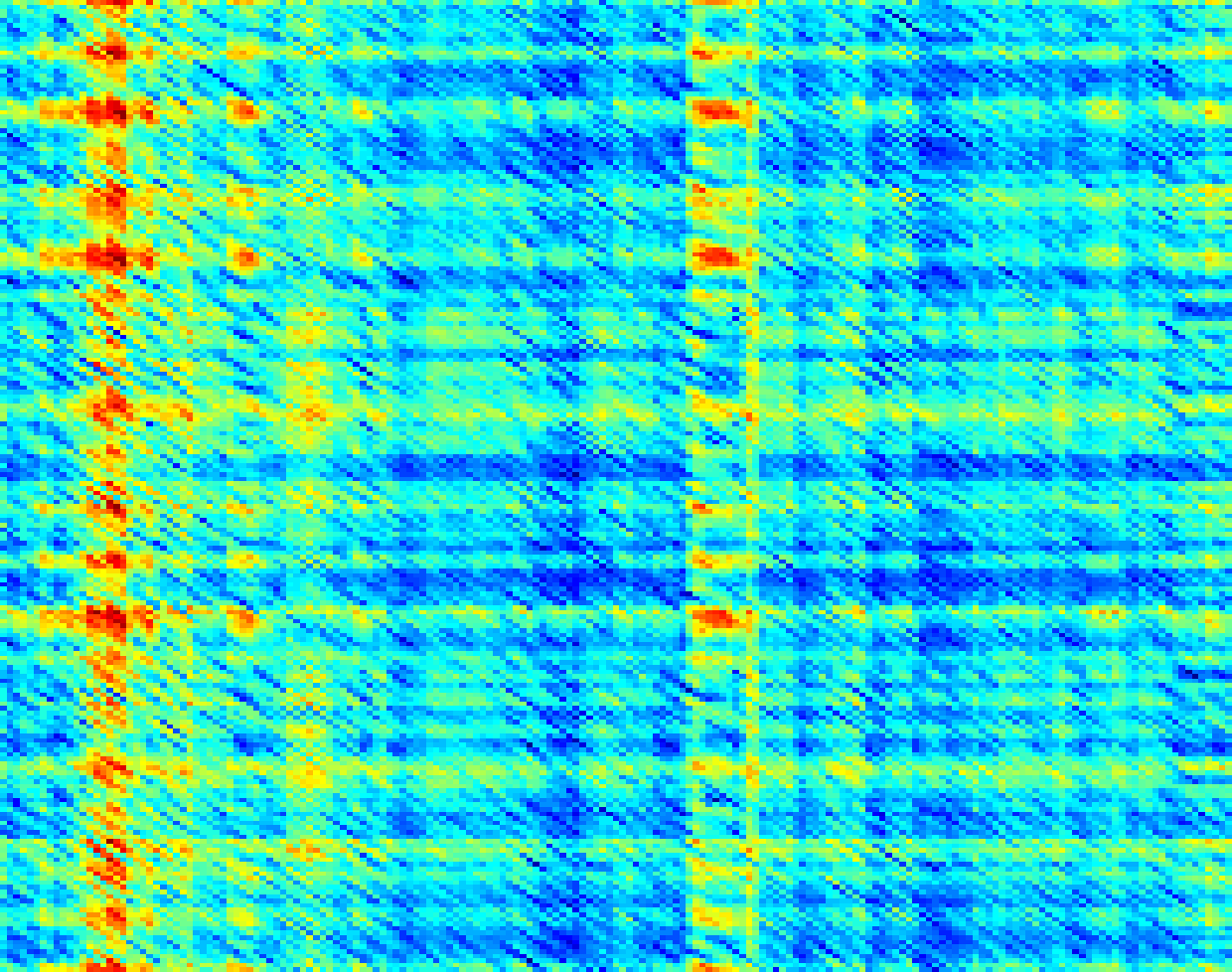} }
	\subfigure[$212 \times 185$ Binary cross-similarity matrix ($B^M$) with $\kappa = 0.05$]{  \includegraphics[width=0.46\columnwidth]{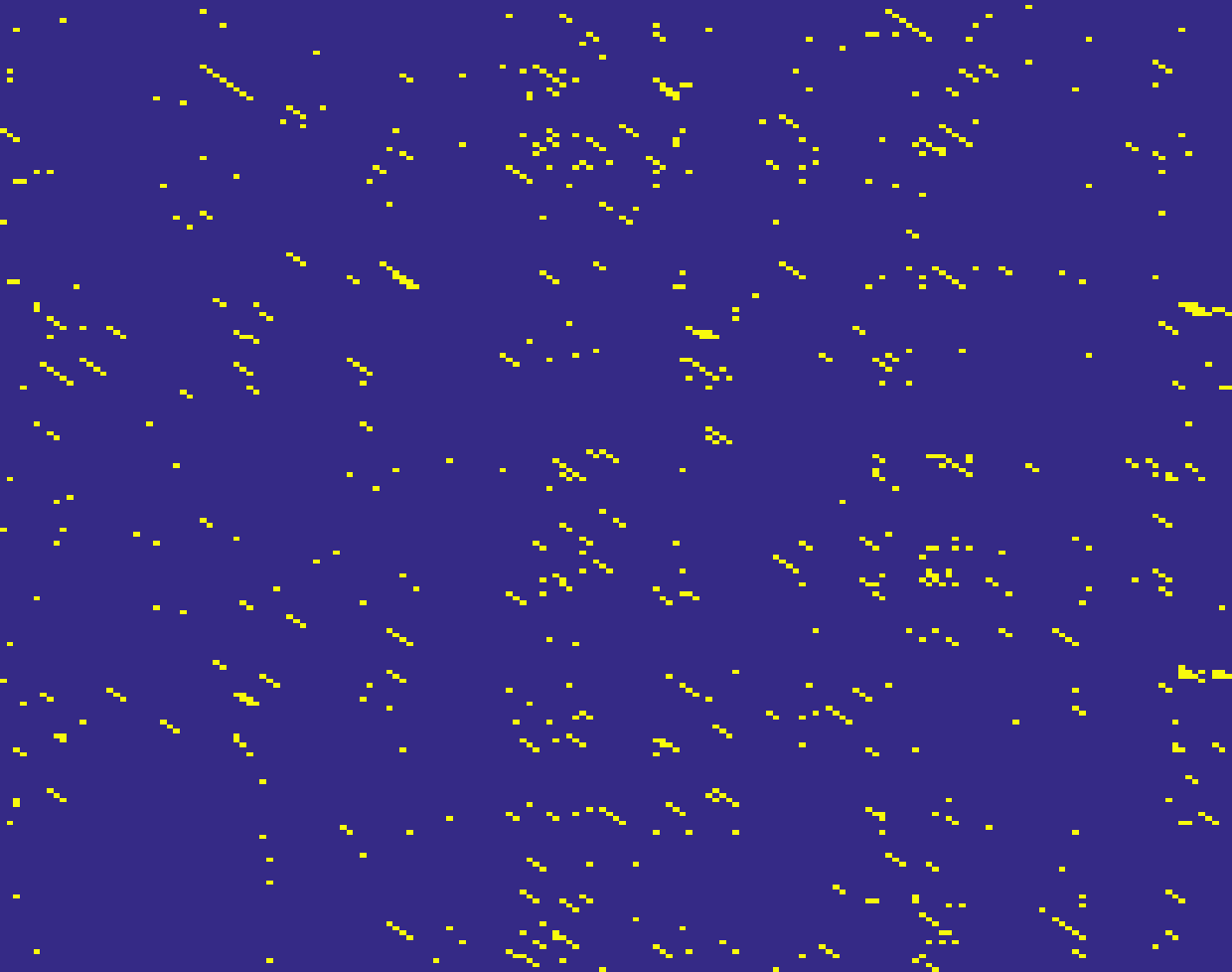} }
	\subfigure[Smith Waterman with local constraints: Score 8 ]{  \includegraphics[width=0.46\columnwidth]{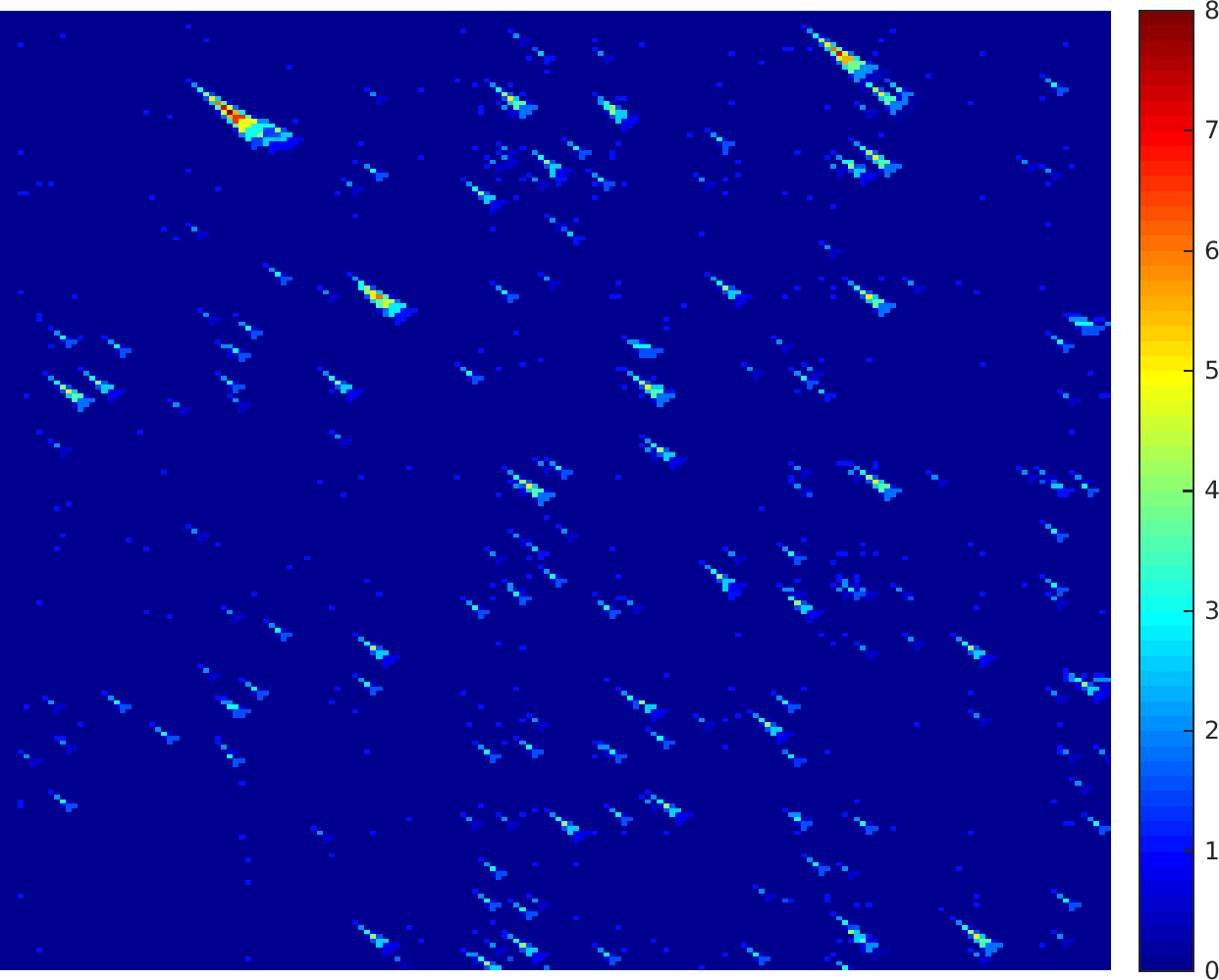} }
	\caption{Cross-similarity matrix and Smith Waterman on MFCC-based SSMs for two songs that are not covers of each other: ``We Can Work It Out" by The Beatles and ``Yesterday" by En Vogue.}
	\label{fig:CrossSimilarityMisMatch}
\end{figure}

Once we have the $B^M$ matrix, we can feed it to the Smith Waterman algorithm, which finds the best local alignment between the two songs, allowing for time shifting and gaps.  Local alignment is a more appropriate choice than global alignment for the cover songs problem, since it is possible that different versions of the same song may have intros, outros, or bridge sections that were not present in the original song, but otherwise there are many sections in common.  We choose a version of Smith Waterman with diagonal constraints, which was shown to work well for aligning binary cross-similarity matrices for chroma in cover song identification \cite{serra2008chroma}.  In particular, we recursively compute a matrix $D$ so that
\begin{equation}
D_{ij} = max \left\{\begin{array}{c} D_{i-1, j-1} + (2 \delta(B_{i-1, j-1}) - 1) + \\ \Delta(B_{i-2, j-2}, B_{i-1, j-1}), \\ \\
D_{i-2, j-1} + (2 \delta(B_{i-1, j-1}) - 1) + \\ \Delta(B_{i-3, j-2}, B_{i-1, j-1}), \\ \\
D_{i-1, j-2} + (2 \delta(B_{i-1, j-1}) - 1) + \\ \Delta(B_{i-2, j-3}, B_{i-1, j-1}), \\ \\
0
\end{array} \right\}
\end{equation}
where $\delta$ is the Kronecker delta function and 
\begin{equation}
\Delta(a, b) = \left\{\begin{array}{cc} $0$ & $b = 1$ \\ $-0.5$ & $b = 0$,  $a = 1$ \\ $-0.7$ & $b = 0$, $a = 0$  \end{array} \right\}
\end{equation}
\begin{figure}
	\centering
	\includegraphics[width=0.3\columnwidth]{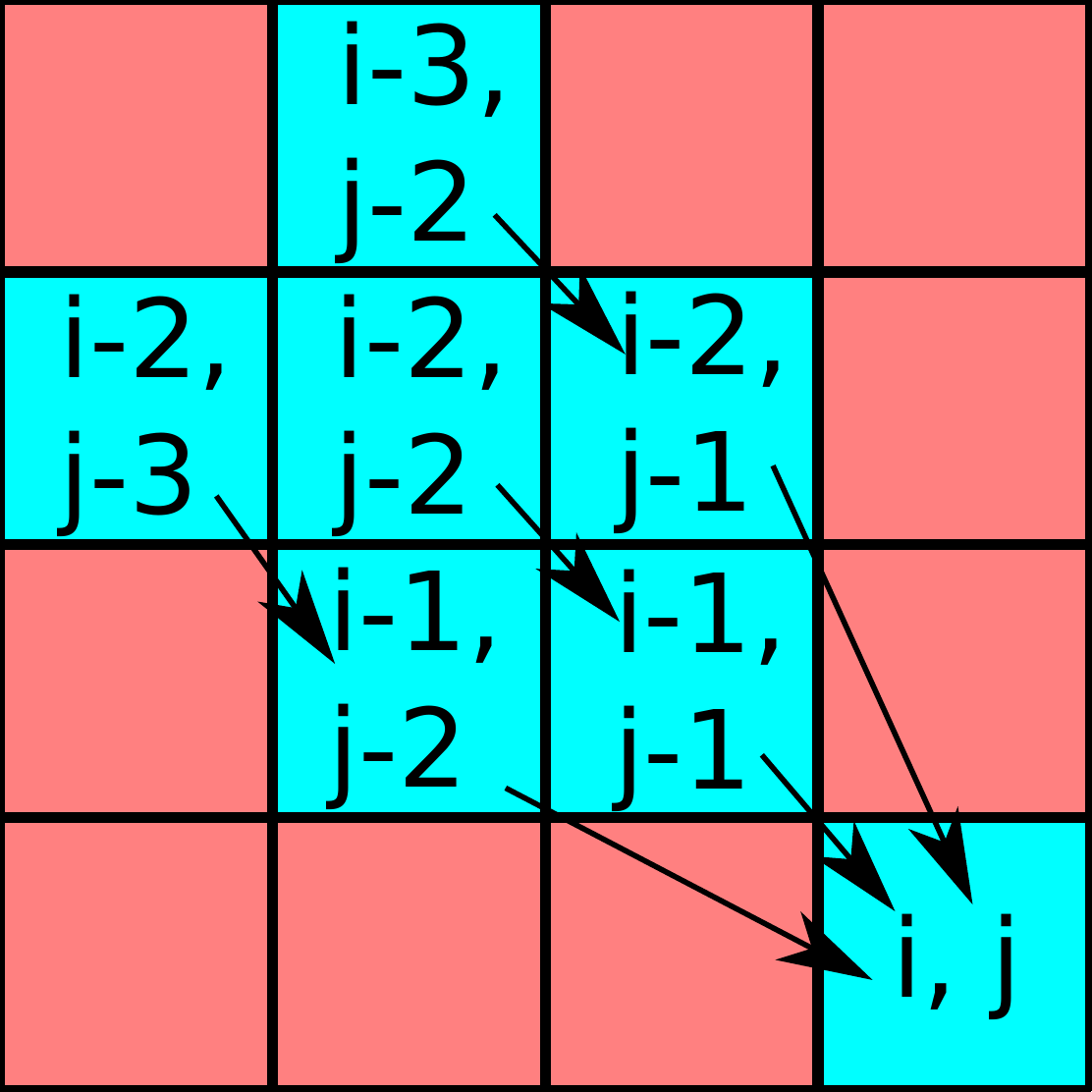}%
	\caption{Constrained local matching paths considered in Smith Waterman, as prescribed by \cite{serra2008chroma}.}
	\label{fig:SWPaths}
\end{figure}
The $(2 \delta(B_{i-1, j-1}) - 1)$ term in each line is such that there will be a +1 score for a match and a -1 score for a mismatch.  The $\Delta$ function is the so-called ``affine gap penalty" which gives a score of $-0.5 - 0.7(g-1)$ for a gap of length $g$.  The local constraints are to bias Smith Waterman to choosing paths along near-diagonals of $B^M$.  This is important since in musical applications, we do not expect large gaps in time in one song that are not in the other, which would show up as horizontal or vertical paths through the $B^M$ matrix.  Rather, we prefer gaps that occur nearly simultaneously in time for a poorly matched beat or set of beats in an otherwise well-matching section.  Figure~\ref{fig:SWPaths} shows a visual representation of the paths considered through $B^M$.

Figure~\ref{fig:CrossSimilarityMatch} shows an example of a CSM, $B^M$, and resulting Smith Waterman for a true cover song pair.  Several long diagonals are visible, indicating large chunks of the two songs are in correspondence, and this gives rise to a large score of $93.1$ between the two songs.  Figure~\ref{fig:CrossSimilarityMisMatch} shows the CSM, $B$, and Smith Waterman for two songs which are not versions of each other.  By contrast, there are no long diagonals, and this pair only receives a score of 8.

\section{Results}
\label{sec:Results}

\subsection{Covers 80}
\label{sec:covers80}
To benchmark our algorithm, we apply it to the standard ``Covers 80" dataset \cite{ellis2007covers80}, which consists of 80 sets of two versions of the same song, most of which are pop songs from the past three decades.  There are designated two sets of songs A and B, each with exactly one version of every pair.  To benchmark our algorithm on this dataset, we follow the scheme in \cite{ellis2006identifying} and \cite{ellis2007}.  That is, given a song from set A, compute the Smith Waterman score from all songs from set B and declare the cover song to be the one with the maximum score.  Note that a random classifier would only get 1/80 in this scheme.  The best scores reported on this dataset are 72/80 \cite{ravuri2010cover}, using a support vector machine on several different chroma-derived features.

Table~\ref{tab:MFCCScores} shows the correctly identified songs based on the maximum score, given variations of the parameters we have in our algorithm.  We achieve a maximum score of $42/80$ for a variety of parameter combinations.  The nearest neighbor fraction $\kappa$ and the dimension of the SSM image have very little effect, but increasing the number of beats per block has a positive effect on the performance.  The stability of $\kappa$ and $d$ are encouraging from a robustness standpoint, and the positive effect increasing the number of beats per block suggests that the shape of medium scale musical expressions are more discriminative than smaller ones.


\begin{table}[htbp]
\caption{The number of songs that are correctly ranked as the most similar in the Covers 80 dataset, varying paramters.  $\kappa$ is the nearest neighbor fraction, $B$ is the number of beats per block, and $d$ is the resized dimension of the Euclidean Self-Similarity images.}
\begin{tabular}{|l|r|r|r|r|}
\hline
Kappa = 0.05 & \multicolumn{1}{l|}{B = 8} & \multicolumn{1}{l|}{B = 10} & \multicolumn{1}{l|}{B = 12} & \multicolumn{1}{l|}{B = 14} \\ \hline
d = 100 & 30 & 33 & 36 & 40 \\ \hline
d = 200 & 31 & 33 & 36 & 39 \\ \hline
d = 300 & 31 & 34 & 36 & 40 \\ \hline
Kappa = 0.1 & \multicolumn{1}{l|}{B = 8} & \multicolumn{1}{l|}{B = 10} & \multicolumn{1}{l|}{B = 12} & \multicolumn{1}{l|}{B = 14} \\ \hline
d = 100 & 35 & 39 & 41 & 42 \\ \hline
d = 200 & 36 & 38 & 42 & 42 \\ \hline
d = 300 & 36 & 38 & 41 & 41 \\ \hline
Kappa = 0.15 & \multicolumn{1}{l|}{B = 8} & \multicolumn{1}{l|}{B = 10} & \multicolumn{1}{l|}{B = 12} & \multicolumn{1}{l|}{B = 14} \\ \hline
d = 100 & 36 & 42 & 41 & 42 \\ \hline
d = 200 & 36 & 41 & 41 & 42 \\ \hline
d = 300 & 38 & 42 & 42 & 41 \\ \hline
\end{tabular}
\label{tab:MFCCScores}
\end{table}

\begin{figure}
	\centering
	\subfigure[Shape-based timbre]{  \includegraphics[width=0.47\columnwidth]{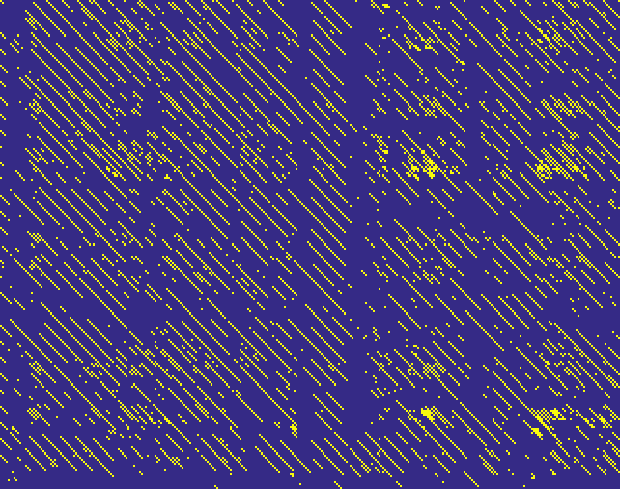} }
	\subfigure[Chroma delay embedding]{  \includegraphics[width=0.47\columnwidth]{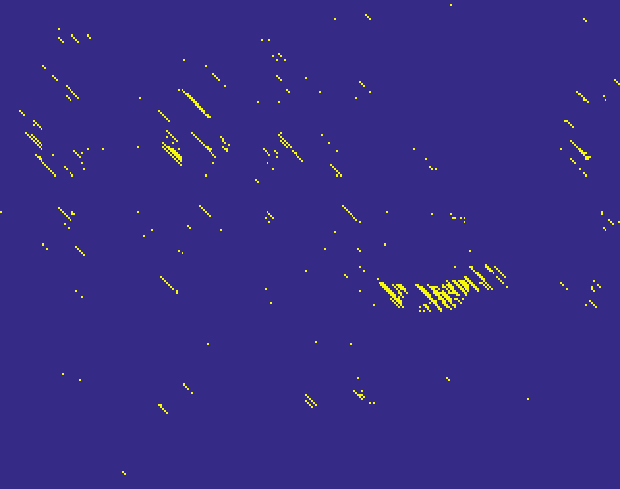} }
	\caption{Corresponding portions of the binary cross-similarity matrix between Marvin Gaye's ``Got To Give It Up" and Robin Thicke's ``Blurred Lines" for both shape-based timbre (our technique) and chroma delay embedding}
	\label{fig:BlurredLines}
\end{figure}

In addition to the Covers 80 benchmark, we apply our cover songs score to a recent popular music controversy, the ``Blurred Lines" controversy \cite{miao2013blurred}.  Marvin Gaye's estate argues that Robin Thicke's recent pop song ``Blurred Lines" is a copyright infringement of Gaye's ``Got To Give It Up."  Though the note sequences differ between the two songs, ruling out any chance of a high chroma-based score, Robin Thicke has said that his song was meant to ``evoke an era" (Marvin Gaye's era) and that he derived significant inspiration from ``Got To Give It Up" specifically \cite{miao2013blurred}.  Without making a statement about any legal implications, we note that our timbral shape-based score between ``Blurred Lines" and ``Got To Give It Up" is in the $99.9^{\text{th}}$ percentile of all scores between songs in group A and group B in the Covers 80 dataset, for $\kappa = 0.1$, $B = 14$, and $d = 200$.  Unsurprisingly, when comparing ``Blurred Lines" with all other songs in the Covers 80 database plus ``Got To Give It Up," ``Got To Give It Up" was the highest ranked.  For reference, binary cross similarity matrices are shown in Figure~\ref{fig:BlurredLines}, both for our timbre shape based technique and the delay embedding chroma technique in \cite{serra2009cross}.  The timbre-based cross-similarity matrix is densely populated with diagonals, while the pitch-based one is not.

\section{Conclusions And Future Work}

We show that timbral information in the form of MFCC can indeed be used for cover song identification.  Most prior approaches have used Chroma-based features averaged over intervals.  By contrast, we show that an analysis of the fine relative shape of MFCC features over intervals is another way to achieve good performance.  This opens up the possibility for MFCC to be used in much more flexible music information retrieval scenarios than traditional audio fingerprinting.  

On the more technical side, we should note that for comparing shape, L2 of SSMs for cross-similarity is fairly simple and not robust to local re-parameterizations in time between versions, though we tried many other isometry invariant shape descriptors that were significantly slower and yielded inferior performance in initial implementation.  In particular, we tried curvature descriptors (ratio of arc length to chord length), Gromov-Hausdorff distance after fractional iterative closest points aligning MFCC block curves \cite{phillips2007outlier}, and Earth Mover's distance between SSMs \cite{shirdhonkar2008approximate}.  If we are able to find another shape descriptor which performs better than our current scheme but is slower, we may still be able to make it computationally feasible by using the ``Generalized Patch Match" algorithm \cite{barnes2010generalized} to reduce the number of pairwise block comparisons needed by exploiting coherence in time.  This is similar in spirit to the approximate nearest neighbors schemes proposed in \cite{tavenard2012efficient} for large scale cover song identification, and we could adapt their sparse Smith Waterman algorithm to our problem.  In an initial implementation of generalized patch match for our current scheme, we found we only needed to query about 15\% of the block pairs.


\section{Supplementary Material}
\label{sec:GUI}
We have documented our code and uploaded directions for performing all experiments run in this paper.  We also created an open source graphical user interface which can be used to interactively view cross-similarity matrices and to examine the shape of blocks of audio after 3D PCA using OpenGL.  All code can be found in
the ISMIR2015 directory at 

\url{github.com/ctralie/
         PublicationsCode}.

\section{Acknowledgements}

Chris Tralie was supported under NSF-DMS 1045133 and an NSF Graduate Fellowship. Paul Bendich was supported by NSF 144749.  
John Harer and Guillermo Sapiro are thanked for valuable feedback. The authors would also like to thank the Information Initative at Duke (iiD) for stimulating this collaboration.

\bibliographystyle{plain}
\bibliography{CoverSongsArXiV}

\begin{thebibliography}{10}

\bibitem{barnes2010generalized}
Connelly Barnes, Eli Shechtman, Dan~B Goldman, and Adam Finkelstein.
\newblock The generalized patchmatch correspondence algorithm.
\newblock In {\em Computer Vision--ECCV 2010}, pages 29--43. Springer, 2010.

\bibitem{bello2007audio}
Juan~Pablo Bello.
\newblock Audio-based cover song retrieval using approximate chord sequences:
  Testing shifts, gaps, swaps and beats.
\newblock In {\em ISMIR}, volume~7, pages 239--244, 2007.

\bibitem{Bogert1963}
Bruce~P Bogert, Michael~JR Healy, and John~W Tukey.
\newblock The quefrency alanysis of time series for echoes: Cepstrum,
  pseudo-autocovariance, cross-cepstrum and saphe cracking.
\newblock In {\em Proceedings of the symposium on time series analysis},
  volume~15, pages 209--243. chapter, 1963.

\bibitem{casey2006importance}
Michael Casey and Malcolm Slaney.
\newblock The importance of sequences in musical similarity.
\newblock In {\em Acoustics, Speech and Signal Processing, 2006. ICASSP 2006
  Proceedings. 2006 IEEE International Conference on}, volume~5, pages V--V.
  IEEE, 2006.

\bibitem{ellis2006identifying}
Daniel~PW Ellis.
\newblock Identifying'cover songs' with beat-synchronous chroma features.
\newblock {\em MIREX 2006}, pages 1--4, 2006.

\bibitem{ellis2007beat}
Daniel~PW Ellis.
\newblock Beat tracking by dynamic programming.
\newblock {\em Journal of New Music Research}, 36(1):51--60, 2007.

\bibitem{ellis2007covers80}
Daniel~PW Ellis.
\newblock The ``covers80" cover song data set.
\newblock {\em URL: http://labrosa. ee. columbia.
  edu/projects/coversongs/covers80}, 2007.

\bibitem{ellis2007}
Daniel~PW Ellis and Courtenay~Valentine Cotton.
\newblock The 2007 labrosa cover song detection system.
\newblock {\em MIREX 2007}, 2007.

\bibitem{ellis2012large}
Daniel~PW Ellis and Bertin-Mahieux Thierry.
\newblock Large-scale cover song recognition using the 2d fourier transform
  magnitude.
\newblock In {\em The 13th international society for music information
  retrieval conference}, pages 241--246, 2012.

\bibitem{foote2000automatic}
Jonathan Foote.
\newblock Automatic audio segmentation using a measure of audio novelty.
\newblock In {\em Multimedia and Expo, 2000. ICME 2000. 2000 IEEE International
  Conference on}, volume~1, pages 452--455. IEEE, 2000.

\bibitem{foucard2010multimodal}
R{\'e}mi Foucard, J-L Durrieu, Mathieu Lagrange, and Ga{\"e}l Richard.
\newblock Multimodal similarity between musical streams for cover version
  detection.
\newblock In {\em Acoustics Speech and Signal Processing (ICASSP), 2010 IEEE
  International Conference on}, pages 5514--5517. IEEE, 2010.

\bibitem{humphrey2013data}
Eric~J Humphrey, Oriol Nieto, and Juan~Pablo Bello.
\newblock Data driven and discriminative projections for large-scale cover song
  identification.
\newblock In {\em ISMIR}, pages 149--154, 2013.

\bibitem{junejo2008cross}
Imran~N Junejo, Emilie Dexter, Ivan Laptev, and Patrick P{\'e}rez.
\newblock Cross-view action recognition from temporal self-similarities.
\newblock In {\em Proceedings of the 10th European Conference on Computer
  Vision: Part II}, pages 293--306. Springer-Verlag, 2008.

\bibitem{kim2008music}
Samuel Kim, Erdem Unal, and Shrikanth Narayanan.
\newblock Music fingerprint extraction for classical music cover song
  identification.
\newblock In {\em Multimedia and Expo, 2008 IEEE International Conference on},
  pages 1261--1264. IEEE, 2008.

\bibitem{mcfee2014analyzing}
Brian McFee and Daniel~PW Ellis.
\newblock Analyzing song structure with spectral clustering.
\newblock In {\em 15th International Society for Music Information Retrieval
  (ISMIR) Conference}, 2014.

\bibitem{miao2013blurred}
Emily Miao and Nicole~E Grimm.
\newblock The blurred lines of what constitutes copyright infringement of
  music: Robin thicke v. marvin gaye’s estate.
\newblock {\em WESTLAW J. INTELLECTUAL PROP.}, 20:1, 2013.

\bibitem{nieto2014music}
Oriol Nieto and Juan~Pablo Bello.
\newblock Music segment similarity using 2d-fourier magnitude coefficients.
\newblock In {\em Acoustics, Speech and Signal Processing (ICASSP), 2014 IEEE
  International Conference on}, pages 664--668. IEEE, 2014.

\bibitem{perea2013sliding}
Jose~A Perea and John Harer.
\newblock Sliding windows and persistence: An application of topological
  methods to signal analysis.
\newblock {\em Foundations of Computational Mathematics}, pages 1--40, 2013.

\bibitem{phillips2007outlier}
Jeff~M Phillips, Ran Liu, and Carlo Tomasi.
\newblock Outlier robust icp for minimizing fractional rmsd.
\newblock In {\em 3-D Digital Imaging and Modeling, 2007. 3DIM'07. Sixth
  International Conference on}, pages 427--434. IEEE, 2007.

\bibitem{ravuri2010cover}
Suman Ravuri and Daniel~PW Ellis.
\newblock Cover song detection: from high scores to general classification.
\newblock In {\em Acoustics Speech and Signal Processing (ICASSP), 2010 IEEE
  International Conference on}, pages 65--68. IEEE, 2010.

\bibitem{salamon2012melody}
Justin Salamon, Joan Serr{\`a}, and Emilia G{\'o}mez.
\newblock Melody, bass line, and harmony representations for music version
  identification.
\newblock In {\em Proceedings of the 21st international conference companion on
  World Wide Web}, pages 887--894. ACM, 2012.

\bibitem{serra2007music}
J~Serra.
\newblock Music similarity based on sequences of descriptors: tonal features
  applied to audio cover song identification.
\newblock {\em Department of Information and Communication Technologies,
  Universitat Pompeu Fabra, Barcelona, Spain}, 2007.

\bibitem{serra2008chroma}
Joan Serra, Emilia G{\'o}mez, Perfecto Herrera, and Xavier Serra.
\newblock Chroma binary similarity and local alignment applied to cover song
  identification.
\newblock {\em Audio, Speech, and Language Processing, IEEE Transactions on},
  16(6):1138--1151, 2008.

\bibitem{serra2012unsupervised}
Joan Serra, Meinard M{\"u}ller, Peter Grosche, and Josep~Lluis Arcos.
\newblock Unsupervised detection of music boundaries by time series structure
  features.
\newblock In {\em Twenty-Sixth AAAI Conference on Artificial Intelligence},
  2012.

\bibitem{serra2009cross}
Joan Serra, Xavier Serra, and Ralph~G Andrzejak.
\newblock Cross recurrence quantification for cover song identification.
\newblock {\em New Journal of Physics}, 11(9):093017, 2009.

\bibitem{shirdhonkar2008approximate}
Sameer Shirdhonkar and David~W Jacobs.
\newblock Approximate earth mover’s distance in linear time.
\newblock In {\em Computer Vision and Pattern Recognition, 2008. CVPR 2008.
  IEEE Conference on}, pages 1--8. IEEE, 2008.

\bibitem{smith1981identification}
Temple~F Smith and Michael~S Waterman.
\newblock Identification of common molecular subsequences.
\newblock {\em Journal of molecular biology}, 147(1):195--197, 1981.

\bibitem{tavenard2012efficient}
Romain Tavenard, Herv{\'e} J{\'e}gou, and Mathieu Lagrange.
\newblock Efficient cover song identification using approximate nearest
  neighbors.
\newblock 2012.

\bibitem{urbano2011melodic}
Juli{\'a}n Urbano, Juan Llor{\'e}ns, Jorge Morato, and Sonia
  S{\'a}nchez-Cuadrado.
\newblock Melodic similarity through shape similarity.
\newblock In {\em Exploring music contents}, pages 338--355. Springer, 2011.

\end{thebibliography}

\end{document}